\documentclass[aps,prb,10pt,longbibliography]{revtex4-2} 

\usepackage[colorlinks, allcolors=blue]{hyperref}

\usepackage{amsmath}  
\usepackage{amsfonts} 
\usepackage{graphicx} 
\begin{document}

\title{Analytical solution of the Sommerfeld-Page equation}

\author{Zurab K. Silagadze}
\email{silagadze@inp.nsk.su} 
\affiliation{Budker Institute of Nuclear Physics, 630 090, Novosibirsk, Russia}
\affiliation{Novosibirsk State University, 630 090, Novosibirsk, Russia}


\begin{abstract}
The Sommerfeld-Page equation describes the non-relativistic dynamics of a classical electron modeled by a sphere of finite size with a uniform surface charge density. It is a delay differential equation, and almost no exact solution of this equation was known until recently. However, progress has been made,  and  an analytical solution was recently found for an almost identical delay differential equation, which arose in the context of the mathematical modeling of COVID-19 epidemics. Inspired by this research, we offer a pedagogical exposition of how one can find an analytical solution of the Sommerfeld-Page equation.
\end{abstract}

\maketitle 

\section{Introduction}
As Philip Pearle remarked, ``The state of  the  classical electron theory reminds one of a house under construction that was abandoned by its  workmen upon receiving news of an approaching plague. The plague in this case, of course, was  quantum theory. As a result, classical electron theory stands with many interesting unsolved or partially solved problems" (p. 213 of Ref. \cite{Pearle1982}). 

One of the oldest problems of classical electron theory is the radiation reaction problem. It goes back to 1881, when J.~J.~Thomson, 18 years before his experimental discovery of the electron, guided by a hydrodynamic analogy, introduced the concept of electromagnetic mass \cite{Dresden1993}. Since then, many eminent scientists including Lorentz, Abraham, Dirac, Born, Feynman and Landau have been working on the radiation reaction problem without a final solution. The relevant literature is vast, and we only mention some of the monographs in this area \cite{Kosyakov_2007,Yaghjian_2006,Rohrlich_2007,spohn_2004,Parrott_1987}  and a few enlightening review articles \cite{Coleman1982,Pearle1982,Jimenez_1999,Erber_1961}.

One might think that this old and out of fashion problem, although annoying, is not relevant today, when we have sophisticated methods at our disposal for solving quantum field theory puzzles that are far more challenging.  However, the subject is interesting in and of itself \cite{Pearle1982}, as the nontrivial combination of classical mechanics and electromagnetism can lead to fascinating and unexpected new insights into old puzzles. 


Second, self-interactions are complex problems to address in quantum theory since they often lead to divergences. Renormalization techniques try to circumvent these issues, but, as Richard Feynman stated in this Nobel Prize lecture, ``the renormalization theory is simply a way to sweep the difficulties of the divergences of electrodynamics under the rug" \cite{Feynman_1966}. In contrast to quantum field theory, many classical gauge theory models can be completely or partially solved, and make the structure of self-actions explicit \cite{Kosyakov_2007}.

Finally, with the development of multipetawatt laser systems, classical and quantum radiation reactions are currently of great practical importance, since they are inevitable in experiments with high-intensity lasers \cite{Blackburn2020,Burby_2020}. For this reason, interest in this topic has increased significantly in recent years \cite{Blackburn2022}. Therefore, although attempts to discuss the structure of the electron solely in the context of classical electrodynamics are certainly pointless, by modernizing these attempts one can learn some interesting lessons and develop important approaches of practical interest.

In our article, we consider one specific aspect of the problem of the radiation reaction: the equation of motion of a uniformly charged spherical shell, considered as a model of a classical electron of finite size. In the next section, we describe this equation, which is a delay differential equation. Since this family of differential equations is little known to most physics students, we give a brief overview of their characteristics in  sections III to V.
We then describe an analytical solution to the homogeneous equation of motion of a uniformly charged spherical shell for some physically interesting initial conditions. This solution was first found in the context of mathematical modeling of the COVID-19 epidemics \cite{Dell’Anna2020}, and we will meet it again in the context of the inhomogeneous equation of motion of the electron in the last section.

it is used also in the section devoted to the consideration of the inhomogeneous case. 

\section{Sommerfeld-Page equation}
The Abraham-Lorentz equation \cite{Griffiths_2023}, which describes the motion of a radiating point particle taking into account the radiation reaction effects, is a third-order equation. As a result, it allows so-called runaway solutions, where the particle spontaneously accelerates in the absence of an external force, and causality-violating so-called pre-acceleration solutions, where the particle accelerates before a force is applied.
The finite-size electron model has been introduced from the very first developments of the radiation reaction theory to avoid non-physical runaway and acausal behavior \cite{Sommerfeld_1905,Page_1918}. If modeled as a uniformly charged spherical shell of radius $R$, then, in the non-relativistic approximation, the equation of motion of an electron submitted to an external force $F(t)$ and taking into account the electromagnetic self-force, has the form \cite{Pearle1982,Erber_1961,Levine_1977,Bohm_1948,Spohn_2011}
\begin{equation}
    m_0\frac{dV}{dt}=F(t)+\frac{e^2}{3R^2c}\left [V\left(t-\frac{2R}{c}\right)-V(t)\right ],
    \label{eq1}
\end{equation}
where $m_0$ is the mechanical (not electromagnetic) mass and $c$ is the speed of light. This equation was implicit in \cite{Sommerfeld_1905} and almost explicit in \cite{Page_1918} (Page's result contains an unfortunate trivial error in the last two lines of his equation (42) \cite{Pearle1982}), which is why it is called the Sommerfeld-Page equation in the literature. To avoid misunderstandings, we emphasize that in equation (\ref{eq1}), the velocity of the center of the sphere $V(t)$ is a function of the argument $t$ (time), and $V(t- 2R/c)$ is the same function of the retarded argument $ t-2R/c$.
 
If we assume that $$\frac{e^2}{3R^2c}\left[V\left(t-\frac{2R}{c}\right)-V(t)\right ]\approx -\frac{2e^2}{3Rc^2}\,\frac{dV}{dt},$$ then Eq.(\ref{eq1}) suggests that $(2e^2)/(3Rc^2)$ can be considered as an electromagnetic contribution to the physical mass of the electron:
\begin{equation}
    m=m_0+\frac{2e^2}{3Rc^2}.
    \label{eq2}
\end{equation}
If we take $m_0=0$ in this equation, we get the classical electron radius as $R_c=(2e^2)/(3mc^2)$, for which the entire mass of the electron is of electromagnetic origin.

It is convenient to introduce the time it takes light to travel the distance $R_c$ \cite{Levine_1977,Griffiths_2010}:
\begin{equation}
    \tau=\frac{R_c}{c}=\frac{2e^2}{3mc^3}\approx 6.2\cdot 10^{-24}~\mathrm{s}.
    \label{eq3}
\end{equation}
Then  $m_0=m-(2e^2)/(3Rc^2)=m(1-c\tau/R)$, and the equation (\ref{eq1}) can be rewritten in the form
\begin{equation}
    \frac{dV}{dt}=\alpha\left[V(t)-V(t-T)\right ]+f(t),
    \label{eq4}
\end{equation}
where 
\begin{equation}
    \alpha=\frac{c^2\tau}{2R^2(c\tau/R-1)},\;\;T=\frac{2R}{c},\;\; f(t)=\frac{F(t)}{m(1-c\tau/R)}.
    \label{eq5}
\end{equation}
Equation (\ref{eq4}) is a delay differential equation.  Such equations are usually not covered in physics courses.  Therefore, we will briefly review some of their properties in the following sections.

In the case of the Sommerfeld-Page equation, the delay of the order of $2R/c$ is caused by the electromagnetic field traversing the electron. Indeed, the total self-force on a charged sphere is obtained by integrating the electromagnetic force that one infinitesimal part of the sphere exerts onto another infinitesimal part over the entire surface of the sphere \cite{Pearle1982}. The somewhat surprising fact that the overall delay can be described by just one constant $2R/c$ is due to the spherical symmetry of the charge distribution \cite{Levine_1977}.

\section{Short history of delay differential equations}
Delay differential equations were first introduced in the 18th century by mathematicians including Bernoulli, Laplace, Poisson, Condorcet, Lacroix, Boole, Combescure and Oltramare \cite{Schmidt1911,Hale_2006}. 
However, the modern history of delay differential equations begins with Volterra's beautiful works on viscoelasticity and the predator-prey models, which were surprisingly largely ignored by his contemporaries and thus had no immediate impact on the development of the field \cite{Hale_1977}. 
Volterra's work on elasticity laid the foundation for the theory of integro-differential equations, which are found in various branches of mathematical physics when considering retarded effects \cite{Whittaker_1941}.  While his contributions to mathematical biology are now widely recognized as having been of decisive importance in stimulating new research in this field \cite{Israel_2002}.

Only after 1940, with the development of ship stabilization and automatic steering systems, for which delay in feedback mechanisms is of paramount importance, did delay differential equations attract attention \cite{Hale_2006,Hale_1977}. In 1830, English mathematician and astronomer George Airy, using a delay differential equation in his attempt to understand how the human voice is produced \cite{Airy_1830,Jenkins_2013}, prophetically noted:``My object is gained if I have called the attention of the Society to a law hitherto (I believe) unnoticed, but not unfruitful in practical applications" \cite{Airy_1830}. Indeed, today delayed effects play an important role in diverse fields such as economics, population dynamics, engineering,  physiology, chemistry, biology, communication and information technologies, mechanics, and theory of partial differential equations of hyperbolic type \cite{Richard_2003,Pinney_1958}.

The use of retarded (or advanced) differential equations in fundamental physics has so far been very limited, since ordinary differential equations and partial differential equations form the basis of almost all fundamental theories. Such theories assume that changes in physical quantities at a certain space-time point are determined by the values of physical quantities in a region infinitely close to this point. We could expect that below a characteristic length, that measures the granularity of space-time, these theories will be no longer valid.  However, space-time granularity is predicted to exist at the Planck scale, which is significantly smaller than any scale that can currently be reached by experiments, and any delays brought on by this level of granularity can therefore be safely ignored.  In any case, incorporating such retarded or advanced time shifts into the fundamental operators that are currently used in mathematical physics is an interesting and non-trivial task \cite{Atiyah_2010}.

\section{Peculiarities of delay differential equations}
To illustrate the differences between ordinary differential equations and delay differential equations, consider the simplest delay differential equation \cite{Bellman_1963,Erneux_2009,Smith_2011}
\begin{equation}
\frac{dy}{dt}=\alpha\,y(t-\tau),
\label{eq6}
\end{equation}
where $\alpha$ and $\tau$ are real constants, and on the right hand side of the equation, the function $y(t)$ is taken with a retarded argument $t-\tau$. In contrast to the similar ordinary differential equation $\frac{dy}{dt}=\alpha y$, which has only exponential solutions $y\sim e^{\alpha t}$, and despite the fact that it only involves the first derivative, Eq.(\ref{eq6}) admits oscillatory solutions $y(t)\sim\sin(\omega t)$ for some values of the angular frequency $\omega$. Indeed, substituting $y(t)=sin{(\omega t)}$ into Eq.(\ref{eq6}), we find that $y(t)=\text{sin}\left(\omega t\right)$ is a solution if the following conditions are met
\begin{equation}
\cos{(\omega \tau)}=0,\;\;\;\; \omega=-\alpha\sin{(\omega \tau)}.
\label{eq7}
\end{equation}
These conditions are satisfied, for example, for $\alpha\tau=-\frac{\pi}{2}$ and $\omega=\frac{\pi}{2\tau}$.

In fact, the differential equation (\ref{eq6}) is not of first, but of infinite order, since the Taylor expansion of $y(t-\tau)$ contains derivatives of arbitrary order. Thus,
the existence of oscillatory solutions is not surprising. The characteristic equation that follows from (\ref{eq6}) is a transcendental equation
\begin{equation}
\lambda=\alpha e^{-\lambda \tau},
\label{eq8}
\end{equation}
with an infinite number of complex solutions
\begin{equation}
\lambda_k=\frac{1}{\tau}W_k(\alpha\tau),\;\;k=0,\pm 1,\pm 2,\ldots ,
\label{eq9}
\end{equation}
where $W_k$ are branches of the Lambert function \cite{Corless1996,Dence2013}. The principal branch $W_0$ has a Taylor series expansion \cite{Dence2013}
\begin{equation}
W_0(x)=\sum\limits_{n=1}^\infty \frac{(-n)^{n-1}}{n!}\;x^n\approx x-x^2+\frac{3}{2}x^3-\frac{8}{3}x^4+\cdots.
\label{eq10}
\end{equation}
Therefore, for $\tau\to 0$, from (\ref{eq9}) and (\ref{eq10}) we obtain $\lambda_0=\alpha$ and recover the exponential solution of the ordinary differential equation.

The properties of the Lambert function and their generalization in the complex domain are important in the analysis of the stability of differential equations with delay, because if the $\lambda_k$-solution of the characteristic equation has a positive real part, then the corresponding solution grows exponentially \cite{Corless1996}.

A boundary condition is the second significant difference between delay differential equations and ordinary differential equations. The solution of a first-order ordinary differential equation is fully determined by stating the value of $y(t)$ at one point. On the contrary, in order to identify one specific solution of a first-order delay differential equation, its values must be specified for the whole $\tau$-interval. For example, if we are interested in finding $y(t)$ that is a solution to Eq.(\ref{eq6}) for all $t\ge \tau$ and is continuous for $t\ge 0$, we can use as a boundary condition
\begin{equation}
y(t)=f(t),\;\; \mathrm{if}\;\; 0\le t\le \tau,
\label{eq11}
\end{equation}
with $f(t)$ as a known function. For any real, continuous function $f(t)$, the boundary condition Eq.(\ref{eq11}) guarantees the existence of a unique solution of Eq.(\ref{eq6}) in the domain $t\ge \tau$ which is continuous for $t\ge 0$.

\section{Solving delay differential equations: the method of steps}
The elegant and simple method of steps is one of the primary techniques for solving differential equations with delay. This approach demonstrates that an equation has a solution and provides a process to compute it \cite{Bellman_1963}. 

To illustrate this method, consider equation (\ref{eq6}) with $\tau=1$, $\alpha=1$ together with the boundary condition (\ref{eq11}) with a constant function $f(t)=1$. If $1\le t\le 2 $, then we  have $y(t-1)=1$, which implies $\frac{dy}{dt}=1$ in this interval. Hence, integrating $\frac{dy}{dt}=1$ over the interval $[1,t]$, we get 
\begin{equation}
y(t)=t=y(1)+(t-1)=1+(t-1),\;\; \mathrm{if}\;\; 1\le t\le 2.
\label{eq12}
\end{equation}
Similarly, in the interval  $2\le t\le 3$, taking into account (\ref{eq12}), we have
$$\frac{dy}{dt}=y(t-1)=1+(t-2),$$
and integrating over the interval $[2,t]$, we end with
\begin{equation}
y(t)=t+\frac{1}{2}(t-2)^2=1+(t-1)+\frac{1}{2}(t-2)^2,\;\; \mathrm{if}\;\; 2\le t\le 3.
\label{eq13}
\end{equation}
We can continue this process for as long as we like, extending the computation of $y(t)$ step by step from one interval to another. By induction, we can prove the following general result \cite{Bellman_1963}
\begin{equation}
y(t)=\sum\limits_{m=0}^n\frac{(t-m)^m}{m!},\;\; \mathrm{if}\;\; n\le t\le n+1,\;\;n=0,1,2,\ldots.
\label{eq14}
\end{equation}

The solution (\ref{eq14}) is continuous for $t>0$, but its first derivative jumps at $t=1$ and is continuous only for $t>1$. The second derivative jumps at $t=2$ but re-establishes continuity at $t>2$; this regularity persists for higher derivatives too. In this case, the initial discontinuity in the derivative smooths out over time as the initial discontinuity in the derivative is successively propagated to higher order derivatives. However, for other delay differential equations, this does not always happen, and, when solving numerically, one should pay attention to the location of discontinuities in the derivatives \cite{Wille1992}.

\section{Analytical solution of the homogeneous Sommerfeld-Page equation}
Now that we have introduced delay differential equations, let us come back to the physical problem of the Sommerfeld-Page equation.
If $V=0$ at $t<0$, and if at $t=0$ the particle experiences a sudden kick so that $f(t)=V_0\delta(t)$, then integrating Eq.(\ref{eq4}) on the time interval $-\epsilon<t<\epsilon$ and taking the limit $\epsilon\to 0$, we get $V(0)=V_0$. Therefore, to find $V(t)$ for $t>0$ we can consider the homogeneous equation
\begin{equation}
\frac{dV(t)}{dt}=\alpha\left [V(t)-V(t-T)\right ],
\label{eq15}
\end{equation}
with boundary conditions
\begin{equation}
    V(0)=V_0,\;\;\;V(t)=0,\;\;\mathrm{if}\;\;t<0.
    \label{eq16}
\end{equation}
This problem admits an analytical solution \cite{Dell’Anna2020} which can be found by the method of steps in the following way.

When $t<T$, $V(t-T)=0$ and Eq.(\ref{eq15}) becomes an ordinary differential equation with $V(t)=V_0e^{\alpha t}$ as the solution. Let us then assume that we have found $V(t)$ for $t\le nT$, where $n$ is a positive integer. Since $V(t-T)$ is a known function for $nT\le t \le (n+1)T$, Eq.(\ref{eq15}) can be viewed as an inhomogeneous ordinary linear differential equation, and its solution can be obtained by the method of variation of parameters. Namely, for the solution of the form $V(t)=A(t)e^{\alpha t}$, equation (\ref{eq15}) implies that $A(t)$ must satisfy the equation
\begin{equation}
\frac{dA}{dt}=-\alpha e^{-\alpha t}V(t-T).
\label{eq17}
\end{equation}
We can integrate Eq.(\ref{eq17}) over the interval $[nT,t]$ and get 
\begin{equation}
A(t)-A(nT)=-\alpha\int\limits_{nT}^t V(\tau-T)e^{-\alpha \tau}d\tau.
\label{eq18}
\end{equation}
Therefore, after the change of variables $t=nT+t^\prime$, $0\le t^\prime<T$, $\tau=nT+s$, we obtain
\begin{equation} 
V(nT+t^\prime)=e^{\alpha t^\prime} \left ( e^{\alpha nT}A(nT)-\alpha\int\limits_0^{t^\prime} V\left ((n-1)T+s\right )\,e^{-\alpha s}ds\right )=
e^{\alpha t^\prime} \left (V(nT)-\alpha\int\limits_0^{t^\prime} V\left ((n-1)T+s\right )\,e^{-\alpha s}ds\right ),
\label{eq19}
\end{equation}
where the final step relies on the continuity of $V(t)$ at $t=nT$. Equation (\ref{eq19}) implies via iteration
\begin{equation} 
V(nT+t^\prime)=e^{\alpha t^\prime}\left [V(nT)-\alpha V((n-1)T)\int\limits_0^{t^\prime} ds+\ldots +(-\alpha)^n V(0)\int\limits_0^{t^\prime} ds_1\int\limits_0^{s_1} ds_2\int\limits_0^{s_2} ds_3\cdots \int\limits_0^{s_{n-1}} ds_n\right ].
\label{eq20}
\end{equation}
The iteration comes to an end  because $V(s-T)=0$ for $s\le t^\prime <T$. The repeated integrals in the above equation can be calculated by using Cauchy's formula for repeated integration
\cite{Oldam_2006}
\begin{equation}
\int\limits_0^{x} dx_1\int\limits_0^{x_1} dx_2\int\limits_0^{x_2} dx_3\cdots \int\limits_0^{x_{n-1}} f(x_n) dx_n=\frac{1}{(n-1)!}\int\limits_0^x (x-y)^{n-1}f(y)dy,
\label{eq21}
\end{equation}
which gives
\begin{equation}
\int\limits_0^{t^\prime} ds_1\int\limits_0^{s_1} ds_2\int\limits_0^{s_2} ds_3\cdots \int\limits_0^{s_{n-1}} ds_n=\frac{t^{\prime\,n}}{n!}.
\label{eq22}
\end{equation}
Therefore,
\begin{equation}
V(nT+t^\prime)=e^{\alpha t^\prime}\sum\limits_{m=0}^n\frac{(-\alpha t^\prime )^m}{m!}\,V\left((n-m)T\right).
\label{eq23}
\end{equation}
By continuity, in the limit $t^\prime\to T$, we obtain
\begin{equation}
V((n+1)T)=e^{\alpha T}\sum\limits_{m=0}^n\frac{(-\alpha T)^m}{m!}\,V\left((n-m)T\right),
\label{eq24}
\end{equation}
which is a recurrence relation for calculating $V(nT)$ in conjunction with $V(0)=V_0$. The first few values are
\begin{eqnarray} &&
V(T)=V_0e^{\alpha T},\;\;\;V(2T)=V_0\left[e^{2\alpha T}-\alpha Te^{\alpha T}\right ], \;\;\;
V(3T)=V_0\left [e^{3\alpha T}-2\alpha Te^{2\alpha T}+\frac{(-\alpha T)^2}{2!}e^{\alpha T}\right ],
\nonumber \\ &&
V(4T)=V_0\left [e^{4\alpha T}-3\alpha Te^{3\alpha T}+\frac{(-2\alpha T)^2}{2!}e^{2\alpha T}+\frac{(-\alpha T)^3}{3!}e^{\alpha T}\right ].
\label{eq25}
\end{eqnarray}
The following general expression for $V(nT)$ is suggested by these equations \cite{Dell’Anna2020}:
\begin{equation}
V(nT)=V_0\sum\limits_{m=1}^n\frac{(-m\alpha T)^{n-m}}{(n-m)!}\,e^{m\alpha T},\;\;n\ge 1.
\label{eq26}
\end{equation}
We will use induction to prove Eq.(\ref{eq26}). Suppose that for any positive integers less than or equal to $n$, equation (\ref{eq26}) holds true. Then the recursion relation Eq.(\ref{eq24}) implies
\begin{equation}
V((n+1)T)=V_0\left [\frac{(-\alpha T)^n}{n!}\,e^{\alpha T}+\sum\limits_{m_1=0}^{n-1}\sum\limits_{m_2=1}^{n-m_1}\frac{(-\alpha T)^{n-m_2}\,m_2^{n-m_1-m_2}}{m_1!\,(n-m_1-m_2)!}\,e^{(m_2+1)\alpha T}\right ].
\label{eq27}
\end{equation}
We can change the order of summations in (\ref{eq27}) and get
\begin{equation}
V((n+1)T)=V_0\left [\frac{(-\alpha T)^n}{n!}\,e^{\alpha T}+\sum\limits_{m_2=1}^n (-\alpha T)^{n-m_2}\,e^{(m_2+1)\alpha T}\sum\limits_{m_1=0}^{n-m_2}\frac{m_2^{n-m_2-m_1}}{m_1!\,(n-m_2-m_1)!}\right ].
\label{eq28}
\end{equation}
However, according to the binomial theorem
\begin{equation}
\sum\limits_{m_1=0}^{n-m_2}\frac{m_2^{n-m_2-m_1}}{m_1!\,(n-m_2-m_1)!}=\frac{1}{(n-m_2)!}\sum\limits_{m_1=0}^{n-m_2}\frac{(n-m_2)!}{m_1!\,(n-m_2-m_1)!}\,m_2^{n-m_2-m_1}=\frac{(m_2+1)^{n-m_2}}{(n-m_2)!},
\label{eq29}
\end{equation}
and Eq.(\ref{eq28}) becomes
\begin{eqnarray} &&
V((n+1)T)=V_0\left [\frac{(-\alpha T)^n}{n!}\,e^{\alpha T}+\sum\limits_{m_2=1}^n \frac{[-(m_2+1)\alpha T]^{n-m_2}}{(n-m_2)!}\,e^{(m_2+1)\alpha T}\right ]= \nonumber \\ &&
V_0\left [\frac{(-\alpha T)^n}{n!}\,e^{\alpha T}+\sum\limits_{m=2}^{n+1} \frac{(-m\alpha T)^{n+1-m}}{(n+1-m)!}\,e^{m\alpha T}\right ]=
V_0\,\sum\limits_{m=1}^{n+1} \frac{(-m\alpha T)^{n+1-m}}{(n+1-m)!}\,e^{m\alpha T}.
\label{eq30}
\end{eqnarray}
This proves that $V((n+1)T)$ has the form of Eq. (\ref{eq26}).

Equation (\ref{eq26}) makes it obvious that the only crucial variable on which $V(nT)$ depends is 
\begin{equation}
{\cal R}_0=\alpha T=\frac{c\tau/R}{c\tau/R-1}.
\label{eq31}
\end{equation}
This parameter is negative if $c\tau<R$. Since in this case all the exponential terms vanish for large times there are no runaway solutions. This result was emphasized already in \cite{Levine_1977,Griffiths_2010}. In the opposite case $c\tau>R$, ${\cal R}_0>1$ and $V(t)$ grows exponentially with time.

It is interesting to note that when equation (\ref{eq15}) with $\alpha>0$ is used to model the initial spread of the COVID-19 epidemics, the parameter ${\cal R}_0=\alpha T$, also known as the basic reproduction number, represents the average number of cases brought on by one infectious case over the course of the infectiousness period $T$ \cite{Dell’Anna2020}. If ${\cal R}_0>1$, then the total number $F(t)$ \footnote{$F(t)$ replaces $V(t)$ in the delay equation (\ref{eq15}).} of people affected by the epidemics (still infected at time $t$ plus those who have recovered or died by this time) grows exponentially. Therefore, to contain epidemics, it is extremely important that ${\cal R}_0<1$, so that $F(t)$ flattens in time, which indicates a stop in the spread of the epidemic. When ${\cal R}_0=1$, the growth of $F(t)$ with time is essentially linear. This behavior of $F(t)$ can be verified by numerical calculations, which can be based on Eq.(\ref{eq23}) together with either Eq.(\ref{eq24}) or Eq.(\ref{eq26}). For $\alpha>0$, it is easy enough to get reliable results only for times $t<nT$ with $n\sim 20$, as precision is lost when subtracting two very large numbers in alternating sums, whose individual terms become very large for large $n$.

It is remarkable and amusing that the delay equation (\ref{eq15}) can be applied to a completely different situation, the initial spread of COVID-19 epidemics, which hardly has anything
to do with the self-force on a charged sphere in classical electrodynamics. This may look as a serendipitous coincidence, another manifestation of ``the unreasonable effectiveness of mathematics" \cite{Wigner_1960}.  However, it is possible that in this case this "unreasonable effectiveness" has deeper roots. The delay differential equation describing a spread of epidemics (including the logistic term $F(t)/N$, $N$ being the total population, which can neglected during the initial spread of epidemics when $F(t)\ll N$)
$$\frac{dF(t)}{dt}=\alpha\left [F(t)-F(t-T)\right ]\left(1-\frac{F(t)}{N}\right),$$
is known to be linked to non-Markovian dynamics \cite{Dell’Anna2020,Kiss_2015}. It is noteworthy that the radiation reaction problem was also reexamined recently from the perspective of non-Markovian dynamics, and it has been argued that pathologic solutions such as runaway solutions arise from taking the Markovian limit improperly \cite{Hsiang:2022}.

\section{Solution of the inhomogeneous Sommerfeld-Page equation}
Assuming that the external force is known for $t\ge 0$ and for $-T\le t<0$ the ``initial" velocity $V_0(t)$ is a given function, consider the inhomogeneous Sommerfeld-Page equation (\ref{eq4}) with the following boundary condition
\begin{equation}
V(t)=V_0(t),\;\;\mathrm{if}\;\;-T\le t<0.
\label{eq32}
\end{equation}
The need to specify the initial velocity in a finite time interval is equivalent to specifying the entire infinite number of derivatives of the velocity at the initial moment of time and reflects the fact that when deriving equation (\ref{eq1}), an infinite number of degrees of freedom of the electromagnetic field were eliminated \cite{Levine_1977}.

To find the corresponding solution $V(t)$ for $t>0$, it is convenient to introduce a Green function $G(t)$ satisfying the equation 
\begin{equation}
\frac{dG(t)}{dt}-\alpha\left [G(t)-G(t-T)\right ]=\delta(t).
\label{eq33}
\end{equation}
Then it is clear that 
\begin{equation}
V(t)=\int\limits_0^\infty G(t-\tau)f(\tau)d\tau
\label{eq34}
\end{equation}
satisfies (\ref{eq4}) for $t>0$. However, causality requires
\begin{equation}
G(t)=0,\;\;\mathrm{if}\;\;t<0,
\label{eq35}
\end{equation}
and hence $V(t)$ in Eq.(\ref{eq34}) is zero for $t<0$ and does not satisfy the boundary condition (\ref{eq32}). To find a solution that meets this boundary condition, we must add to Eq.(\ref{eq34}) the corresponding homogeneous solution, which can be found as follows \cite{Pearle1982}.

If we change $t$ to $t-\tau$ in equation (\ref{eq33}), multiply both sides by $V_0(\tau)$, then integrate over $\tau$ from $-T$ to $0$, we get
\begin{equation}
\int\limits_{-T}^0V_0(\tau)\frac{dG(t-\tau)}{dt}d\tau-\alpha\int\limits_{-T}^0V_0(\tau)\left[ G(t-\tau)-G(t-\tau-T) \right ]d\tau=\int\limits_{-T}^0V_0(\tau)\delta(t-\tau)d\tau=V_0(t).
\label{eq36}
\end{equation}
But if $-T\le t<0$ and $-T\le \tau\le 0$, then $T+\tau\ge 0$, $t-\tau-T<0$, and according to Eq.(\ref{eq35}) $G(t-\tau-T)=0$. In addition, after integrating by parts, we get
\begin{equation}
\int\limits_{-T}^0V_0(\tau)\frac{dG(t-\tau)}{dt}d\tau=-\int\limits_{-T}^0V_0(\tau)dG(t-\tau)=-\left . V_0(\tau)dG(t-\tau)\right |_{-T}^0+ \int\limits_{-T}^0\frac{dV_0(\tau)}{d\tau} G(t-\tau)d\tau.
\label{eq37}
\end{equation}
When $-T\le t<0$, $G(t-\tau)|_{\tau=0}=0$. Therefore, Eqs.(\ref{eq36}) and (\ref{eq37}) imply
\begin{equation}
V_0(t)=V_0(-T)G(t+T)+\int\limits_{-T}^0\left [ \frac{dV_0(\tau)}{d\tau}-\alpha V_0(\tau)\right ]G(t-\tau)d\tau,\;\;\mathrm{for}\;\; -T\le t<0.
\label{eq38}
\end{equation}
On the other hand, for $t>0$, $-T\le \tau\le 0$, both $t+T$ and $t-\tau$ are non-zero which imply that $G(t+T)$ and $G(t-\tau)$ in the Eq.(\ref{eq38}) satisfy the homogeneous version of Eq.(\ref{eq33}) (without delta-function). Therefore, the right-hand-side of Eq.(\ref{eq38}) is just a solution of the homogeneous Sommerfeld-Page equation we need (with the correct boundary conditions for $-T\le t<0$).

As we see, the solution of the inhomogeneous Sommerfeld-Page equation with the boundary conditions (\ref{eq32}) has the form \cite{Pearle1982}
\begin{equation}
V(t)=V_0(-T)G(t+T)+\int\limits_{-T}^0\left [ \frac{dV_0(\tau)}{d\tau}-\alpha V_0(\tau)\right ]G(t-\tau)d\tau+\int\limits_0^\infty G(t-\tau)f(\tau)d\tau.
\label{eq39}
\end{equation}
It remains to find the Green function $G(t)$. For $t>0$ this function satisfies the homogeneous Sommerfeld-Page equation. If we integrate Eq.(\ref{eq33}) over time over the interval $-\epsilon\le t\le \epsilon$ and then take the limit $\epsilon\to 0$, we get $G(0)=1$. Thus, the initial value problem for $G(t)$ (with boundary conditions $G(t)=0,\;\mathrm{if}\;t<0$, $G(0)=1$) is exactly of the type considered in the previous section, and we can immediately write the solution
\begin{equation}
G(nT+t^\prime)=e^{\alpha t^\prime}\sum\limits_{m=0}^n\frac{(-\alpha t^\prime )^m}{m!}\,G\left((n-m)T\right),
\label{eq40}
\end{equation}
where $0\le t^\prime<T$, and $G(nT)$ is calculated either with the help of the recurrence relation
\begin{equation}
G((n+1)T)=e^{\alpha T}\sum\limits_{m=0}^n\frac{(-\alpha T)^m}{m!}\,G\left((n-m)T\right),
\label{eq41}
\end{equation}
with $G(0)=1$, or directly by using
\begin{equation}
G(nT)=\sum\limits_{m=1}^n\frac{(-m\alpha T)^{n-m}}{(n-m)!}\,e^{m\alpha T},\;\;n\ge 1.
\label{eq42}
\end{equation}
However, Eq.(\ref{eq42}) allows to rewrite Eq.(\ref{eq40}) in a more convenient way. Indeed, these relations imply
\begin{equation}
G(nT+t^\prime)=e^{\alpha t^\prime}\left [\frac{(-\alpha t^\prime)^n}{n!}+\sum\limits_{m_1=0}^{n-1}\frac{(-\alpha t^\prime)^{m_1}}{m_1!}\sum\limits_{m_2=1}^{n-m_1}\frac{(-m_2\alpha T)^{n-m_1-m_2}}{(n-m_1-m_2)!}e^{m_2\alpha T}\right ].
\label{eq43}
\end{equation}
But $\sum\limits_{m_1=0}^{n-1}\sum\limits_{m_2=1}^{n-m_1}$ is the same as $\sum\limits_{m_2=1}^n\sum\limits_{m_1=0}^{n-m_2}$, and after this reversal of the order of summations, the inner sum (in $m_1$) becomes an easily computed binomial sum
\begin{equation}
\sum\limits_{m_1=0}^{n-m_2}\frac{(-\alpha t^\prime)^{m_1}\;(-m_2\alpha T)^{n-m_2-m_1}}{m_1!\;(n-m_2-m_1)!}=\frac{[-\alpha (t^\prime+m_2T)]^{n-m_2}}{(n-m_2)!}.
\label{eq44}
\end{equation}
Therefore,
\begin{equation}
G(nT+t^\prime)=e^{\alpha t^\prime}\left [ \frac{(-\alpha t^\prime)^n}{n!}+\sum\limits_{m_2=1}^n\frac{e^{m_2\alpha T}}{(n-m_2)!}\;[-\alpha(t^\prime+m_2 T)]^{n-m_2} \right ]= \sum\limits_{m=0}^n\frac{e^{\alpha(t^\prime+m_2 T)}}{(n-m)!}\;[-\alpha(t^\prime+m T)]^{n-m}.
\label{eq45}
\end{equation}
Finally, $t^\prime=t-nT$ when $nT\le t<(n+1)T$. Therefore, $t^\prime+mT=t-(n-m)T$ and Eq.(\ref{eq45}) is rewritten as follows
\begin{equation}
G(t)=\sum\limits_{m=0}^n\frac{e^{\alpha(t-(n-m)T)}}{(n-m)!}\;\left [-\alpha(t-(n-m)T)\right ]^{n-m}=   \sum\limits_{m=0}^n\frac{e^{\alpha(t-mT)}}{m!}\;\left [-\alpha(t-mT)\right ]^m,\;\;nT\le t<(n+1)T.
\label{eq46}
\end{equation}
As an example, consider the case of a rectangular force
\begin{equation}
F(t)=\left \{\begin{array}{l} 0,\;\;t\le 0,\\F,\;\; 0<t<2T,\\0,\;\;t\ge 2T,\end{array} \right .
\label{eq47}
\end{equation}
where $F$ is some constant. We will use the same parameters as considered in \cite{Levine_1977,Griffiths_2010}, namely $V_0(t)=0$ and $c\tau/R=2/3$, which imply 
\begin{equation}
T=\frac{2R}{c}=3\tau,\;{\cal R}_0=\alpha T=\frac{c\tau/R}{c\tau/R-1}=-2,\;\alpha=\frac{{\cal R}_0}{T}=-\frac{2}{3\tau},\
\label{eq48a}
\end{equation}
and
\begin{equation}
f(t)=\frac{F(t)}{m\left (1-c\tau/R\right )}=\left \{\begin{array}{l} 0,\;\;t\le 0,\\\frac{3F}{m},\;\; 0<t<2T,\\0,\;\;t\ge 2T.\end{array} \right .
\label{eq48}
\end{equation}
Using Eq.(\ref{eq46}), we get
\begin{equation}
G(t)=\left \{\begin{array}{l} 0,\;\;t\le 0,\\e^{\alpha t},\;\; 0\le t<T,\\e^{\alpha t}-\alpha(t-T)e^{\alpha(t-T)},\;\;T\le t< 2T,\\e^{\alpha t}-\alpha(t-T)e^{\alpha(t-T)}+\frac{\alpha^2}{2}(t-2T)^2e^{\alpha(t-2T)},\;\;2T\le t< 3T.\end{array} \right .
\label{eq49}
\end{equation}
On the other hand, from
\begin{equation}
V(t)=\int\limits_0^\infty G(t-\tau)f(\tau)d\tau=\frac{3F}{m}\int\limits_0^{2T} G(t-\tau)d\tau=
\frac{3F}{m}\int\limits_{t-2T}^tG(s)ds,
\label{eq50}
\end{equation}
we get
\begin{equation}
a(t)=\frac{dV}{dt}=\frac{3F}{m}\left [G(t)-G(t-2T)\right ],
\label{eq51}
\end{equation}
and using Eq.(\ref{eq49}) we find
\begin{equation}
a(t)=\frac{3F}{m}e^{\alpha t}\times\left \{\begin{array}{l} 0,\;\;t\le 0,\\1,\;\; 0\le t<T,\\1-\alpha(t-T)e^{-\alpha T},\;\;T\le t< 2T,\\1-\alpha(t-T)e^{-\alpha T}+\left [\frac{\alpha^2}{2}(t-2T)^2-1\right ]e^{-2\alpha T},\;\;2T\le t< 3T.\end{array} \right .
\label{eq52}
\end{equation}
It is easy to check that for $\alpha=-2/(3\tau)$, $T=3\tau$ this result coincides with the result given by equation (46) in \cite{Griffiths_2010}.

\section{Concluding remarks}
The expression for the Green's function (\ref{eq46}), combined with Eq.(\ref{eq39}), provides a solution to the Sommerfeld-Page equation (\ref{eq4}) for the general boundary condition (\ref{eq32}). As a consequence, the Sommerfeld-Page equation does not have pathological (runaway) solutions for $R>c\tau$, but such solutions are an inevitable feature of this equation for $R<c\tau$. 
Therefore, if a charged sphere is used as a model of an electron, and the radius of the sphere
is greater than the classical radius of the electron, no runaway or other pathological solutions arise

What about the singular limit $R\to c\tau$? When $\alpha<0$, then for any integer $m$ we have $-\alpha\int\limits_0^\infty e^{\alpha x}(-\alpha x)^mdx=m!$ and  Eq.(\ref{eq46}) indicates that there exists a well-defined limit for $\alpha G(t)$ when $\alpha\to -\infty$ (that is when $R\to (c\tau)_+$). Namely,
\begin{equation}
\lim_{\alpha\to -\infty}\alpha G(t)=-\sum\limits_{n=0}^\infty \delta(t-nT).
\label{eq53}
\end{equation}
This result was obtained in \cite{Pearle1982} by the Laplace transform method (note that the Green's function defined in \cite{Pearle1982} is not $G(t)$ but $G(t)/[m(1-c\tau/R)]$).

If we expand $V(t-2R/c)-V(t)$ up to the second order in the small (compared to $t$) parameter $2R/c$, then from Eq.(\ref{eq1}) we get the Abraham-Lorentz equation \cite{Griffiths_2023,Pearle1982}
\begin{equation}
m\frac{dV}{dt}=F(t)+\frac{2e^2}{3c^3}\,\frac{d^2 V}{dt^2}=F(t)+m\tau\frac{da}{dt},\;\;a=\frac{dV}{dt},
\label{eq54}
\end{equation}
which is known to have unphysical runaway solutions. If we consider Eq.(\ref{eq54}) as a mere approximation of Eq.(\ref{eq1}), this is hardly surprising, since equation (\ref{eq1}) itself has runaway solutions, when $R<c\tau$. Based on physical intuition, Landau and Lifshitz suggested \cite{Landau:1975} a way to avoid runaway solutions and replace Eq.(\ref{eq54}) by a physically more relevant equation that can be obtained from Eq.(\ref{eq54}) by replacing the acceleration $a$ with its zeroth-order approximation $a=F/m$:
\begin{equation}
m\frac{dV}{dt}=F(t)+\tau\frac{dF}{dt}.
\label{eq55}
\end{equation}
A rigorous justification of Landau and Lifshitz's proposal,  which should discuss the concepts of the geometric theory of singular perturbations, in particular reduction to a so-called slow manifold \cite{Bel_1982,Spohn_2000,Chicone_2001,Chicone:2001mw,Burby_2020}, is beyond the scope of this article. In what follows, we confine ourselves to some general remarks on these questions.

The fundamental equations of physics, such as Maxwell's equations of electrodynamics or Einstein's equations of general relativity, are of the hyperbolic type and therefore entail finite propagation velocities of classical fields. As a result, for example, only after a delay that corresponds to the time the light signal takes to propagate from one charge to another is the perturbation of one charge felt by the other, so that the force acting on one charge depends not only on the instantaneous positions of the other charges, but also on the past history of the system of charges \cite{Kibble_2004}. As a consequence, it is expected that physical phenomena with delayed interactions should be modeled by functional differential equations, in which the delay, generally speaking, is of a functional type. Such functional differential equations are extremely complicated and hard to analyze. For this reason, in order to obtain useful dynamical predictions from realistic models, it is common to expand the functional differential equation in terms of small delays and truncate the expansion to a finite order, thus replacing the functional differential equations of motion with approximations that are ordinary (or partial) differential equations \cite{Bel_1982,Chicone_2001,Chicone:2001mw}. However, this truncation can, and usually does, result in unphysical runaway modes, which may just be artifacts of the truncation. Therefore, additional efforts are needed to obtain physically reasonable approximations and distinguish between physical solutions that are asymptotically attracted to the so-called slow manifold and non-physical runaway solutions that are asymptotically repelled from the slow manifold \cite{Bel_1982,Chicone_2001,Chicone:2001mw}.

Since all calculations can be performed analytically, we believe that the results of this paper support Ref.\cite{Levine_1977}'s assertion that the non-relativistic motion of a charged shell is of particular pedagogical value. A more general case of the Sommerfeld-Page equation, in the presence of an external magnetic field was studied in \cite{Spohn_2011}, and the stability and long time behavior of the solutions were investigated.

Last but not least, this problem allows students to become familiar with delay differential equations. The relativistic generalization of the Sommerfeld-Page equation was developed by Caldirola \cite{Caldirola1956}. His approach was based on the existence of the smallest interval of proper time, the so-called chronon, and this granularity in time leads naturally to differential equations with delay. The idea of elemental space or time intervals is a recurring theme in scientific literature \cite{Esposito_2010} and is currently an active area of research \cite{Hossenfelder_2013}. It is not excluded that differential equations with delay will play an increasingly important role in fundamental physics \cite{Atiyah_2010}.

\section{Conflict of interest statement}
The author has no conflicts to disclose.

\bibliography{SP_Equation}

\end{document}